\documentclass[12pt]{iopart}
\usepackage[english]{babel}
\usepackage{latexsym}
\usepackage{setstack}
\usepackage{iopams}
\usepackage{amscd}
\newcommand{\bdm}{\begin{displaymath}}
\newcommand{\edm}{\end{displaymath}}
\newcommand{\bdn}{\begin{eqnarray}}
\newcommand{\edn}{\end{eqnarray}}
\newcommand{\bay}{\begin{array}{c}}
\newcommand{\eay}{\end{array}}
\newcommand{\ben}{\begin{enumerate}}
\newcommand{\een}{\end{enumerate}}
\newcommand{\beq}{\begin{equation}}
\newcommand{\eeq}{\end{equation}}
\newtheorem{lem}{Lemma}[section]
\newtheorem{teo}{Theorem}[section]
\newtheorem{pro}{Proposition}[section]

\eqnobysec

\begin{document}

\title[Decay of a bound state under a time-periodic perturbation]{Decay of a bound state under a time-periodic perturbation: a toy case}
\author{Michele Correggi\dag, Gianfausto Dell'Antonio\ddag}
\address{\dag International School for Advanced Studies SISSA/ISAS, Trieste, Italy}
\address{\ddag Centro Linceo Interdisciplinare\footnote[3]{On leave from Dipartimento di Matematica, Universit\`{a} di Roma, ``La Sapienza'', Italy}, Roma, Italy}
\eads{\mailto{correggi@sissa.it}, \mailto{gianfa@sissa.it}}

\begin{abstract}
	We study the time evolution of a three dimensional quantum particle, initially in a bound state, under the action of a time-periodic zero range interaction with ``strength'' \( \alpha(t) \). Under very weak generic conditions on the Fourier coefficients of \( \alpha(t) \), we prove complete ionization as \( t \rightarrow \infty \). We prove also that, under the same conditions, all the states of the system are scattering states.
\end{abstract}

\section{Introduction}

In this paper we shall study the asymptotically complete ionization of a system given by a quantum particle interacting with a time-dependent singular potential in three dimensions. The Hamiltonian of the system is formally
\bdm
	H(t) = H_0 + H_I(t)
\edm
where \( H_0 \) is a zero range perturbation at the origin of the Laplacian and \( H_I(t) \) is heuristically given by \( \alpha(t) \delta(\bi{x} -\bi{r}) \) where \( \bi{r} \in \mathbb{R}^3 \setminus \{0\} \) and \( \alpha(t) \) is a periodic function with period \( T \).
\newline
This kind of models have been widely studied (see e.g. \cite{Corr1,Corr2,Cost1,Cost2,Cost3,Cost4,Cost5}) as toy models of more complicated physical problems, such as strong laser ionization of Rydberg atoms or dissociation of molecules. Indeed time-dependent point interactions are an interesting example of time-dependent perturbations that are not small in any sense with respect to the unperturbed Hamiltonian, so that time-dependent perturbation theory (and therefore Fermi's golden rule) can not be applied. On the other hand, since such models are solvable, namely all the spectral and scattering data can be explicitly calculated, the problem of asymptotically complete ionization can be studied in a non-pertubative way. Indeed one can explicitly prove that, starting at time \( t = 0 \) from a bound state \( \varphi \) of the system, the survival probability
\bdm
	|\theta(t)|^2 = \Big| \Big( \varphi, \: U(t,0) \varphi \Big) \Big|^2
\edm
has a power law decay to zero as \( t \rightarrow \infty \) (see \cite{Corr2,Cost1} and references therein).
\newline
Essentially using Laplace transform techniques (for a review of the methods used, we shall refer to \cite{Corr2}), we shall prove that the system shows asymptotically complete ionization under suitable generic conditions on the Fourier coefficients of \( \alpha(t) \) and that the survival probability has a power law decay for large time.
\newline
We stress the non-perturbative nature of the result. Indeed the complete ionization does not depend on the size of \( \alpha(t) \) and it holds even if \( \alpha(t) \) is very big (so that the time-dependent perturbation is small - in the sense of quadratic forms - with respect to the unperturbed Hamiltonian) or very small (so that the perturbation is large) or fast oscillating. Moreover the asymptotic behavior is independent of the period \( T \). In particular there is asymptotically complete ionization, even if the period is very large, as for time-adiabatic perturbations.
\newline  
In Section 2 we shall introduce the model, the equations for the ``charges'' and their Laplace transforms, which will be the main objects under investigation. Applying analytic Fredholm theorem to such equations, in Sections 3,4 and 5 we shall identify the singularities of their solutions on the closed right half plane; in Section 6 we shall derive the main results about ionization.

\section{The model}

The model we are going to study describes a quantum particle subjected to a time-dependent zero range interaction. In absence of the time-periodic perturbation, the Hamiltonian describes a zero range interaction placed at the origin and of strength \( - 1/ 4 \pi \). This system has a bound state of energy \( -1 \) and normalized eigenfunction
	\beq
	\label{Initial}
		\Psi_0 (\bi{x}) = \frac{e^{-|\bi{x}|}}{\sqrt{4\pi} |\bi{x}|}
	\eeq
The remaining part of the spectrum is absolutely continuous and coincides with \( \mathbb{R}^+ \). 
\newline
The time-dependent perturbation is a zero range interaction placed at a point \( \bi{r} \neq 0 \) and with time-periodic strength \( \alpha(t) \) with period \( T \).
\newline
The entire system is then described (see \cite{Albe1}) by the time-dependent self-adjoint Hamiltonian \( H(t) \),
	\bdm
		\mathcal{D}(H(t)) = \bigg\{ \Psi \in L^2(\mathbb{R}^3) \: \big| \: \exists \: q^{(1)}(t), \: q^{(2)}(t) \in \mathbb{C}, 
	\edm 
	\beq 
		\varphi_{\lambda}(\bi{x}) \equiv \bigg(\Psi(\bi{x}) -  q^{(1)}(t) \: \mathcal{G}_{\lambda}(\bi{x}) - q^{(2)}(t) \: \mathcal{G}_{\lambda}(\bi{x}- \bi{r}) \bigg) \in H^2(\mathbb{R}^3) \bigg\}
	\eeq
	\beq
	\label{Operator}
		\big( H(t) + \lambda \big) \Psi(\bi{x}) = \big( - \Delta + \lambda \big) \varphi_{\lambda}(\bi{x})
	\eeq
where \( \lambda \) is an arbitrary positive parameter. The ``charges'' \( q^{(i)}(t) \) are determined by the boundary conditions
	\beq
	\label{Boundary}
		\begin{array}{l}
			\varphi_{\lambda}(0) =  -\displaystyle{\frac{1 + i \lambda}{4 \pi}} \: q^{(1)}(t) - \mathcal{G}_{\lambda}(\bi{r}) \: q^{(2)}(t)	\\
			\mbox{}	\\
			\varphi_{\lambda}(\bi{r}) = \displaystyle{\frac{4 \pi \alpha(t) - i \lambda}{4 \pi}} \: q^{(2)}(t) - \mathcal{G}_{\lambda}(\bi{r}) \: q^{(1)}(t)	\\
		\eay
	\eeq
and
	\bdm
		\mathcal{G}_{\lambda}(\bi{x} - \bi{x}^{\prime}) = \frac{e^{-\sqrt{\lambda}|\bi{x} - \bi{x}^{\prime}|}}{4\pi |\bi{x}-\bi{x}^{\prime}|}
	\edm
is the Green function of the Laplacian.
\newline
It is well known (see \cite{Dell1,Dell2,Figa1,Saya1,Yafa1}) that the solution of the time-dependent Schr\"{o}dinger equation
	\beq
	\label{Schro}
		i \frac{\partial \Psi_t}{\partial t} = H(t) \Psi_t
	\eeq
associated to the operator (\ref{Operator}) is given by 
	\bdm
		\Psi_t(\bi{x}) = U_0(t-s) \Psi_s (\bi{x}) + i \int_s^t d \tau \: \bigg[ \: q^{(1)}(\tau) \: U_0(t-\tau ; \bi{x}) +
	\edm
	\beq
	\label{State}
		+ \: q^{(2)}(\tau) \: U_0(t-\tau ; \bi{x}-\bi{r}) \bigg]
	\eeq
where \( U_0(t) = \exp( i \Delta t) \), \( U_0(t;\bi{x}) \) is the kernel associated to the free propagator and the charges \( q^{(j)}(t) \) satisfy a system of Volterra integral equations for \( t \geq s \),
	\bdm
		q^{(1)}(t) + \frac{\sqrt{-2i}}{\pi} \int_s^t d \tau \: q^{(2)}(\tau) \int_{\tau}^t d\sigma \: \frac{U_0(\sigma-\tau ; \bi{r})}{\sqrt{t- \sigma}} +
	\edm
	\beq
	\label{Equation1}
		- \frac{1}{\sqrt{-\pi i}} \int_s^t d\tau \: \frac{q^{(1)}(\tau)}{\sqrt{t-\tau}} = 4 \sqrt{\pi i} \int_s^t d\tau \: \frac{\big(U_0(\tau) \Psi_s\big)(0)}{\sqrt{t-\tau}}
	\eeq
	\bdm
		q^{(2)}(t) + \frac{\sqrt{-2i}}{\pi} \int_s^t d \tau \: q^{(1)}(\tau) \int_{\tau}^t d\sigma \: \frac{U_0(\sigma-\tau ; \bi{r})}{\sqrt{t- \sigma}} +
	\edm
	\beq
	\label{Equation2}
		+ 4 \sqrt{\pi i} \int_s^t d\tau \: \frac{\alpha(\tau) \: q^{(2)}(\tau)}{\sqrt{t-\tau}} = 4 \sqrt{\pi i} \int_s^t d\tau \: \frac{\big(U_0(\tau) \Psi_s\big)(\bi{r})}{\sqrt{t-\tau}}
	\eeq
We are interested in studying asymptotic complete ionization of system defined by (\ref{Operator}) and (\ref{Schro}), starting by the normalized bound state (\ref{Initial}) at time \( t = 0 \). Moreover we shall require that \( \alpha(t) \) is a real continuous periodic function with period \( T \), so that it can be decomposed in a Fourier series, for each \( t \in \mathbb{R}^+ \), and the series converges uniformly on every compact subset of the real line. More precisely, in terms of Fourier coefficients of \( \alpha(t) \), we assume
	\beq
	\label{Conditions2}
		\begin{array}{ll}
			1.	&	\alpha(t) = \displaystyle{\sum_{n \in \mathbb{Z}}} \: \alpha_n \: e^{-i n \omega t} \: , \: \{ \alpha_n \} \in \ell_1(\mathbb{Z}) 	\\
			\mbox{}	&	\\
			2.	&	\alpha_n = \alpha_{-n}^*	
		\eay
	\eeq	
We now introduce a \emph{generic} condition on \( \alpha(t) \), that will be used later on. Let \( \mathcal{T} \) be the right shift operator on \( \ell_2(\mathbb{N}) \), i.e.
	\beq
		\big( \mathcal{T} \alpha \big)_n \equiv \alpha_{n+1}
	\eeq
we say that \( \alpha = \{ \alpha_n \} \in \ell_2(\mathbb{Z}) \) is \emph{generic} with respect to \( \mathcal{T} \), if \( \tilde{\alpha} \equiv \{ \alpha_n \}_{n>0} \in \ell_2(\mathbb{N}) \) satisfies the following condition
	\beq
	\label{Genericity}
		e_1 = \big( 1,0,0, \ldots \big) \in \overline{\bigvee_{n=0}^{\infty} \mathcal{T}^n \tilde{\alpha}}
	\eeq
For a detailed discussion of genericity condition see \cite{Cost1}. Notice that 
\beq
	\label{alpha}
	\alpha_0 \equiv \frac{1}{T} \int_0^T \alpha(t) dt
\eeq
does not enter in the condition.
\newline
By simple estimates on the sup norm of \( r_j(t) \equiv q^{(j)}(t) \: e^{-bt} \), it is easy to prove that the charges \( q^{(j)}(t) \) have at most an exponential behavior as \( t \rightarrow \infty \), i.e. asymptotically \( |q^{(j)}(t)| \leq A_j e^{b_j t} \). 
\newline
Therefore the Laplace transform of \( q^{(j)}(t) \), denoted by 
	\bdm
		\tilde{q}^{(j)}(p) \equiv \int_0^{\infty} dt \: e^{-pt} q^{(j)}(t)
	\edm
exists ans is analytic at least for \( \Re(p) > b_0 \). Hence, applying the Laplace transform to equations (\ref{Equation1}) and (\ref{Equation2}), one has
	\beq
	\label{Laplace1}
		\tilde{q}^{(1)}(p) =  - \frac{1}{(2 \pi)^{\frac{3}{2}} r} \: \frac{e^{-r \sqrt{-ip}}}{1 - \sqrt{-ip}} \: \tilde{q}^{(2)}(p) + F_1(p)
	\eeq
	\beq
	\label{Laplace2}
		\tilde{q}^{(2)}(p) =  - \frac{4 \pi}{\sqrt{-ip}} \sum_{k \in \mathbb{Z}} \alpha_k \tilde{q}^{(2)}(p+i \omega k) + \frac{e^{-\sqrt{-ip} r}}{2 \pi r \sqrt{-2 \pi i p}} \: \tilde{q}^{(1)}(p) + F_2(p)
	\eeq
where the explicit expression of \( F_i(p) \) for the initial datum (\ref{Initial}) is given by 
	\bdm
		F_1(p) \equiv - \frac{2i \sqrt{2 \pi}}{1 + ip}
	\edm
	\bdm
		F_2(p) \equiv - \frac{2i \sqrt{2 \pi}}{\sqrt{-ip}} \: \frac{e^{-\sqrt{-ip}r} - e^{-r}}{r(1 + ip)}
	\edm
Let us start considering the system of equations (\ref{Laplace1}) and (\ref{Laplace2}), for the specific initial datum (\ref{Initial}): analyticity at least for for \( \Re(p) > b_0 \) suggests to choose the branch cut of the square root along the negative real line: if \( p = \varrho \: e^{i \vartheta} \),
	\beq
	\label{Branch}
		\sqrt{p} = \sqrt{\varrho} \:\: e^{i \vartheta / 2}
	\eeq
with \( -\pi < \vartheta \leq \pi \). 
\newline
Before dealing with the behavior of the solution, let us simplify the problem: setting \( q^{(j)}_n(p) \equiv \tilde{q}^{(j)}(p+i \omega n) \) we obtain a sequence of functions on the strip \( \mathcal{I} = \{ p \in  \mathbb{C}, \: 0 \leq \Im(p) < \omega \} \) and setting \( q_j(p) \equiv \{ q^{(j)}_n(p) \}_{n \in \mathbb{Z}} \), equations (\ref{Laplace1}) and (\ref{Laplace2}) can be rewritten
	\beq
	\label{EqL1}
		q_1(p) = \mathcal{M}_1 \: q_2(p) + G_1(p)
	\eeq
	\beq
	\label{EqL2}
		q_2(p) = \mathcal{L} \: q_2(p) + \mathcal{M}_2 \: q_1(p) + G_2(p)
	\eeq
where
	\beq
	\label{M1Operator}
		\big( \mathcal{M}_1 \: q \big)_n (p) \equiv - \frac{1}{(2\pi)^{\frac{3}{2}} r} \frac{e^{-r\sqrt{\omega n - ip}}}{1 - \sqrt{\omega n - ip}} \:\: q_n(p) 
	\eeq
	\beq
	\label{M2Operator}
		\big( \mathcal{M}_2 \: q \big)_n (p) \equiv \frac{1}{(2\pi)^{\frac{3}{2}} r} \frac{e^{-r\sqrt{\omega n - ip}}}{4 \pi \alpha_0 + \sqrt{\omega n - ip}} \:\: q_n(p) 
	\eeq
	\beq	
	\label{LOperator}
		 \big( \mathcal{L} \: q \big)_n (p) \equiv - \frac{4 \pi}{4 \pi \alpha_0 + \sqrt{\omega n - ip}} \: \underset{k \neq 0}{\sum_{k \in \mathbb{Z}}} \: \alpha_k \: q_{n+k}(p)
	\eeq
and \( G_j(p) = \{ g^{(j)}_n(p) \}_{n \in \mathbb{Z}} \) with
	\beq
		g^{(1)}_n(p) \equiv \frac{2 i \sqrt{2 \pi}}{1 - \omega n + ip}
	\eeq
	\beq
		g^{(2)}_n(p) \equiv - \frac{2 i \sqrt{2 \pi}}{r} \frac{e^{-r\sqrt{\omega n - ip}} - e^{-r}}{(4 \pi \alpha_0 + \sqrt{\omega n - ip})(1- \omega n + ip)}
	\eeq

\section{Analyticity on the (open) right half plane}

Let us extend equations (\ref{EqL1}) and (\ref{EqL2}) on the whole open right half plane: we are going to prove that the solution exists and is analytic for \( \Re(p) > 0 \). 
\newline
Let us start with some preliminary results:
	
	\begin{pro}
	\label{CompactM}
		For \( p \in \mathcal{I} \), \( \Re(p) > 0 \), \( \mathcal{M}_j(p) \) are analytic operator-valued functions and \( \mathcal{M}_j(p) \) are compact operators on \( \ell_2(\mathbb{Z}) \).
	\end{pro}
	
	\emph{Proof:}
		Let us consider only \( \mathcal{M}_1 \), since the argument does apply to \( \mathcal{M}_2 \) too.
		\newline
		The analyticity of the operator is a straightforward consequence of the explicit expression (\ref{M1Operator}). Moreover the operator \( \mathcal{M}_1 (p) \) is a multiplication operator in \( \ell_2(\mathbb{Z}) \) and it is bounded and compact since
		\bdm	
			\left\{ \frac{1}{(2\pi)^{\frac{3}{2}} r} \frac{e^{-r\sqrt{\omega n - ip}}}{1 - \sqrt{\omega n - ip}}  \right\} \in \ell_2(\mathbb{Z})
		\edm
		on the open right half plane: indeed the choice (\ref{Branch}) for the branch cut of the square root implies \( \Re(\sqrt{\omega n - ip}) > 0 \), if \( \Re(p) > 0 \). \opensquare 
	
	\begin{pro}
	\label{CompactL}
		For \( p \in \mathcal{I} \), \( \Re(p) > 0 \), \( \mathcal{L}(p) \) is an analytic operator-valued function and \( \mathcal{L}(p) \) is a compact operator on \( \ell_2(\mathbb{Z}) \).
	\end{pro}

	\emph{Proof:}	
		Analyticity for \( \Re(p) > 0 \) easily follows from the explicit expression of the operator. Moreover \( \mathcal{L}(p) \) can be written
		\bdm
			\mathcal{L}(p) = A(p) \: \underset{k \neq 0}{\sum_{k \in \mathbb{Z}}} \: \alpha_k \: \mathcal{T}^{n+k}
		\edm
		where \( A(p) \) is the operator
		\bdm
			(A \: q)_n (p) \equiv A_n(p) \: q_n(p) =  - \frac{4 \pi q_n(p)}{4 \pi \alpha_0 + \sqrt{\omega n - ip}}
		\edm
		and \( \mathcal{T} \) is the right shift operator on \( \ell_2(\mathbb{Z}) \). Since \( \| \mathcal{T} \| = 1 \), the series converges strongly to a bounded operator. Moreover \( A(p) \) is a compact operator for \( \Re(p) > 0 \): \( A(p) \) is the norm limit of a sequence of finite rank operators, because \( \lim_{n \rightarrow \infty} A_n(p) = 0 \). Hence the result follows e.g. from Theorem VI.12 and VI.13 of \cite{Reed1}. \opensquare

	\begin{lem}
	\label{Imaginary1}
		For each \( r,\omega \in \mathbb{R}^+ \) and for \( \Re(p)>0 \)
		\bdm
			\Im \left[ \sqrt{\omega n - ip} + \frac{1}{(2\pi)^3 r^2} \frac{e^{-2r\sqrt{\omega n - ip}}}{1 - \sqrt{\omega n - ip}} \right] < 0 
		\edm
		\( \forall n \in \mathbb{Z} \).
	\end{lem}
	
	\emph{Proof:}
		First of all we want to stress that the choice (\ref{Branch}) for the branch cut implies that \( \Re(\sqrt{\omega n - ip}) > 0 \) and \( \Im(\sqrt{\omega n - ip}) < 0 \), if \( \Re(p) > 0 \).
		\newline
		Calling \( x \equiv \Re(\sqrt{\omega n - ip}) \), \( y \equiv \Im(\sqrt{\omega n - ip}) \) and
		\bdm
			f_r(x,y) \equiv \Im \left[ x + iy + \frac{1}{(2\pi)^3 r^2} \frac{e^{-2r(x+iy)}}{1 - x -iy} \right]
		\edm
		one has
		\bdm
			\left| \frac{1}{(2\pi)^3 r^2} \frac{e^{-2r(x+iy)}}{1 - x-iy} \right| < \frac{1}{(2\pi)^3 r^2 |y|}
		\edm
		and then \( f_r(x,y) \leq 0 \), if \( |y| \geq [(2\pi)^{3/2} r]^{-1} \). Moreover 
		\bdm
			f_r(x,y) = \frac{ (2\pi)^3 r^2 \big[(1-x)^2+y^2\big] y + e^{-2rx} \big[ y \cos(2ry) - (1-x) \sin(2ry) \big]}{(2\pi)^3 r^2 [(1-x)^2+y^2]} 
		\edm
		and the claim is true if \( x \geq 1 \), since \( \sin(2ry) < 0 \) and \( \cos(2ry) > 0 \), for \( y > - [(2\pi)^{3/2} r]^{-1} \).
		\newline  
		Hence it is sufficient to prove that \( f_r(x,y) < 0 \) on the set 
		\bdm
			R = \left\{ (x,y) \in \mathbb{R}^2 \: | \: x<1, - [(2\pi)^{3/2}r]^{-1} < y < 0 \right\} 
		\edm
		Now set 
		\bdm	
			g_r(x,y) \equiv \frac{(2\pi)^3 r^2 [(1-x)^2+y^2] \: f_r(x,y)}{y} 
		\edm  
		and consider
		\bdm
			\frac{\partial  g_r}{\partial y} = 2 (2\pi)^3 r^2 y + 
		\edm
		\bdm	
			- 2 e^{-2rx} \left[ r \sin(2ry) +  \frac{r (1-x)\cos(2ry)}{y} -  \frac{(1-x)\sin(2ry)}{2y^2} \right]
		\edm
		Since, for \( (x,y) \in R \), \( 2 e^{-2rx} r \sin(2r|y|) < 2 (2\pi)^3 r^2 |y| \) and \( 2 r y \cos(2ry) \leq \sin(2ry) \) the partial derivative of \( g_r \) with respect to \( y \) is always negative in \( R \). Thus
		\bdm
			g_r(x,y) \geq g_r(x,0) > 0
		\edm
		In conclusion \( g_r(x,y) > 0 \) and then \( f_r(x,y) < 0 \), \( \forall (x,y) \in R \). \opensquare
		 
	\begin{pro}
	\label{Analyticity}
		The solutions \( \tilde{q}^{(j)}(p) \), \( j =1,2 \), of (\ref{Laplace1}) and (\ref{Laplace2}) are unique and analytic for \( \Re(p) > 0 \).
	\end{pro}
	
	\emph{Proof:} 
		Since \( G_1(p) \in \ell_2(\mathbb{Z}) \) is analytic on the right half plane and thanks to Proposition \ref{CompactM}, we can substitute (\ref{EqL1}) in (\ref{EqL2}) and consider only the second equation. So that (\ref{EqL2}) now read
		\beq
		\label{Eqr1}
			q_2(p) = \big[ \mathcal{L} + \mathcal{M}_2 \mathcal{M}_1 \big] \: q_2(p) + \mathcal{M}_2 \: G_1(p) + G_2(p)
		\eeq
		Then the key point will be the application of the analytic Fredholm theorem (Theorem VI.14 of \cite{Reed1}) to the operator \( \mathcal{L}^{\prime}(p) \equiv \mathcal{L} + \mathcal{M}_2 \mathcal{M}_1 \), in order to prove that \( (I - \mathcal{L}^{\prime}(p))^{-1} \) exists for \( \Re(p) > 0 \). 
		\newline
		So let us begin with the analysis of the homogeneous equation associated to (\ref{Eqr1}),
		\bdm
			q(p) = \mathcal{L}^{\prime}(p) \:  q(p)
		\edm
		and suppose that there exists a nonzero solution \( Q(p) = \{ Q_n(p) \}_{n \in \mathbb{Z}} \). Multiplying both sides of the equation by \( Q_n^* \) and summing over \( n \in \mathbb{Z} \), we have
		\bdm
			\sum_{n \in \mathbb{Z}} \left[ \sqrt{\omega n - ip} + \frac{1}{(2\pi)^3 r^2} \frac{e^{-2r\sqrt{\omega n - ip}}}{1 - \sqrt{\omega n - ip}} \right] \: |Q_n|^2 =  - 4 \pi \sum_{n,k \in \mathbb{Z}} Q^*_n \: \alpha_{k-n} \: Q_{k}
		\edm
		but, since the right hand side is real, because of condition 2 in (\ref{Conditions2}), it follows that
		\bdm
			\Im \left[ \sum_{n \in \mathbb{Z}} \left( \sqrt{\omega n - ip} + \frac{1}{(2\pi)^3 r^2} \frac{e^{-2r\sqrt{\omega n - ip}}}{1 - \sqrt{\omega n - ip}} \right) \: |Q_n|^2 \right] = 0 
		\edm
		and then, by Lemma \ref{Imaginary1}, \( Q_n = 0 \), \( \forall n \in \mathbb{Z} \).
		\newline
 		Since there is no nonzero solution of the homogeneous equation associated to (\ref{Eqr1}) and \( \mathcal{L} \) is compact on the whole open right half plane, analytic Fredholm theorem applies and the result then easily follows, because \( \mathcal{M}_2 G_1(p) + G_2(p) \in \ell_2(\mathbb{Z}) \) and, for each \( n \in \mathbb{Z} \), \( \big[ \mathcal{M}_2 G_1(p) + G_2(p) \big]_n \) is analytic for \( \Re(p) > 0 \). \opensquare

\section{Behavior on the imaginary axis at \( p \neq 0 \)}

The equation for \( q_2(p) \) can be written
	\beq
	\label{Eqr2}
		\big( 4 \pi \alpha_0 + c_n(p) \big) \: q^{(2)}_n(p) = - 4 \pi \underset{k \neq 0}{\sum_{k \in \mathbb{Z}}} \alpha_k q^{(2)}_{n+k}(p) + f^{(2)}_n(p)
	\eeq
where
	\beq
		c_n(p) \equiv \sqrt{\omega n - ip} + \frac{e^{-2r\sqrt{\omega n - ip}}}{(2 \pi)^3 r^2(1 - \sqrt{\omega n - ip})}
	\eeq
	\beq
		f^{(2)}_n(p) \equiv - \frac{2i \sqrt{2 \pi}}{r (1 - \omega n +ip)} \bigg[\frac{(2 \pi)^{\frac{3}{2}} - 1}{(2 \pi)^{\frac{3}{2}}}  \: e^{-r\sqrt{\omega n -ip}} - e^{-r} \bigg]
	\eeq
and it is clear that the solution may have a pole where
	\bdm
		4 \pi \alpha_0 + \sqrt{\omega n - ip} + \frac{e^{-2r\sqrt{\omega n - ip}}}{(2 \pi)^3 r^2(1 - \sqrt{\omega n - ip})} = 0
	\edm
and that the coefficients of the equation for \( q^{(2)}_0 \) fail to be analytic at \( p =i \): for \( p \in \mathcal{I} \), \( \Re(p) = 0 \), and \( n \in \mathbb{Z} \), the unique solution of \( 1 - \sqrt{\omega n - ip} = 0 \) is \( p = i \), \( n = 0 \).
\newline
In the following we shall see that in fact the solution is analytic on the imaginary axis except at most some singularity at \( p = 0\). Let us start considering the position of the eventual pole:
	
	\begin{lem}
	\label{Imaginary2}
		Assume that \( \alpha_0 \) in (\ref{alpha}) is non negative. Then there exists a unique \( n_0 \in \mathbb{N} \) and a unique \( p_0 \in \mathcal{I} \), \( \Re(p) = 0 \), such that
		\bdm	
			4 \pi \alpha_0 + \sqrt{\omega n_0 - i p_0} + \frac{e^{-2r\sqrt{\omega n_0 - ip_0}}}{(2 \pi)^3 r^2(1 - \sqrt{\omega n_0 - ip_0})} = 0
		\edm
		Moreover \( \forall n \in \mathbb{Z} \), \( n < 0 \) and \( \forall p \in \mathcal{I} \), \( \Re(p) = 0 \),
		\bdm
			\Im \left[ 4 \pi \alpha_0 + \sqrt{\omega n - ip} + \frac{e^{-2r\sqrt{\omega n - ip}}}{(2 \pi)^3 r^2(1 - \sqrt{\omega n - ip})} \right] > 0
		\edm
	\end{lem}

	\emph{Proof:}
		Let us first consider the second statement: on the strip \( \mathcal{I} \) and for \( n < 0 \), \( \sqrt{\omega n - ip} \equiv i \lambda \), with \( \lambda \in \mathbb{R} \), \( \lambda > 0 \). Hence
		\bdm
			\Im \big( c_n(i\lambda) \big) = \frac{(2\pi)^3 r^2 (1 + \lambda^2) \lambda + \lambda \cos(2 r \lambda) - \sin(2 r \lambda)}{ (2\pi)^3 r^2 (1 + \lambda^2)}
		\edm  
		and following the proof of Lemma \ref{Imaginary1}, it can be easily proved that the expression above is positive \( \forall \lambda \in \mathbb{R}^+ \). On the other hand, if \( n \geq 0 \) and \( p \in \mathcal{I} \), \( \Re(p) = 0 \), \( \sqrt{\omega n - ip} = \lambda \), with \( \lambda > 0 \), and, \( \forall r, \omega \in \mathbb{R}^+ \), the equation
		\bdm
			(2 \pi)^3 r^2 (4 \pi \alpha_0 + \lambda) (\lambda - 1) = e^{-2r \lambda} 
		\edm
		has a unique solution for \( \lambda \in \mathbb{R}^+ \). Then, since there exists a unique \( p_0 \in \mathcal{I} \), \( \Re(p_0) = 0 \), such that, for fixed \( \lambda \in \mathbb{R}^+ \), the equation \( p_0 = i (\lambda^2 - \omega n_0) \) is satisfied for some \( n_0 \in \mathbb{N} \), the proof is complete. \opensquare

 	\begin{lem}
	\label{AnalyticityI}		
		Assume that \( \alpha_0 \) in (\ref{alpha}) is non negative and that \( \{ \alpha_n \} \) satisfies (\ref{Conditions2}) and the genericity condition with respect to \( \mathcal{T} \) (\ref{Genericity}). Then the solutions of (\ref{Laplace1}) and (\ref{Laplace2}) are unique and analytic on the imaginary axis for \( p \neq 0, i, p_0 \).
	\end{lem}	
		
	\emph{Proof:}
		Since for \( p \in \mathcal{I} \), \( \Re(p) = 0 \), and \( p \neq 0,i,p_0 \), the coefficients of equation (\ref{EqL1}) and (\ref{EqL2}) are analytic (see Lemma \ref{Imaginary2}) and belong to \( \ell_2(\mathbb{Z}) \) and since the operators \( \mathcal{L}, \mathcal{M}_1 \) and \( \mathcal{M}_2 \) are still compact on the same region, it is sufficient to show that the homogeneous equation associated to (\ref{Eqr2}) has no non zero solution, in order to apply analytic Fredholm theorem.
		\newline
		If \( Q_n \) is such a non zero solution, following the proof of Proposition \ref{Analyticity}, we immediately obtain the condition
		\bdm
			\sum_{n \in \mathbb{Z}} \left[ \sqrt{\omega n - ip} + \frac{1}{(2\pi)^3 r^2} \frac{e^{-2r\sqrt{\omega n - ip}}}{1 - \sqrt{\omega n - ip}} \right] \: |Q_n|^2 \in \mathbb{R}
		\edm
		and then Lemma \ref{Imaginary2} guarantees that \( Q_n = 0 \), \( \forall n < 0 \). Now let \( n_1 \in \mathbb{N} \) be such that \( Q_{n_1} \neq 0 \). For \( n < n_1 \), one has \( \sum_{k = n_1}^{\infty} \alpha_{k-n} Q_{k} = 0 \) or, setting \( k = n_1 -1 + k^{\prime} \), for \( n \geq 0 \), 
		\bdm
			\sum_{k^{\prime}=1}^{\infty} \alpha_{k^{\prime} + n} Q_{n_1-1+k^{\prime}} = 0 
		\edm
		and then, for each \( n \geq 0 \),
		\bdm
			\Big(  Q^{\prime} \: , \mathcal{T}^n \alpha \Big)_{\ell_2(\mathbb{N})} = 0
		\edm
		where \( Q^{\prime}_n = Q^*_{n_1-1+n} \) and \( ( \cdot \: , \: \cdot ) \) stands for the standard scalar product on \( \ell_2(\mathbb{N}) \). Finally the genericity condition (\ref{Genericity}) implies that \( Q^{\prime}_1 = Q^*_{n_1} = 0 \), which is contradiction. Hence \( Q_n = 0 \), \( \forall n \in \mathbb{Z} \). \opensquare

	\begin{pro}
	\label{Poles}
		Assume that \( \alpha_0 \) in (\ref{alpha}) is non negative and that \( \{ \alpha_n \} \) satisfies (\ref{Conditions2}) and the genericity condition with respect to \( \mathcal{T} \) (\ref{Genericity}). Then the solutions of (\ref{Laplace1}) and (\ref{Laplace2}) are unique and analytic on the imaginary axis except at most at \( p = 0 \).
	\end{pro}

	\emph{Proof:}
		In the first part of the proof we are going to consider only the equation (\ref{Eqr2}) for \( q_2(p) \) and we shall extend then the results to \( q_1(p) \).
		\newline
		In order to prove analyticity of the solution we need to analyze the behavior of the solution of (\ref{Eqr2}) in a neighborhood of \( p = p_0 \) (see Lemma \ref{Imaginary2}) and \( p =i \) separately and show that it has no singularity, while, for \( p \in \mathcal{I} \), \( \Re(p) = 0 \), and \( p \neq i, p_0 \), the result follows from Lemma \ref{AnalyticityI}.
		\newline
		Let us look for a solution of (\ref{Eqr2}) of the form (for simplicity we are going to omit the index 2)
		\bdm
			q_n = u_n + v_n q_{n_0}
		\edm
		for \( n \neq n_0 \): \( q_n \) satisfies (\ref{Eqr2}) if and only if \( \{ u_n \} \), \( \{ v_n \} \in \ell_2 ( \mathbb{Z} \setminus \{ n_0 \} ) \) are solutions of 
		\beq
		\label{Equn}
			c_n(p) \: u_n = - 4 \pi \underset{k \neq n_0}{\sum_{k \in \mathbb{Z}}} \alpha_{k-n} u_k + f_n^{(2)}(p)
		\eeq
		\beq
		\label{Eqvn}
			c_n(p) \:  v_{n} =  - 4 \pi \underset{k \neq n_0}{\sum_{k \in \mathbb{Z}}} \alpha_{k-n} v_k - 4 \pi \alpha_{n_0-n}
		\eeq
		Existence of non-zero solutions of the homogeneous equations associated to (\ref{Equn}) and (\ref{Eqvn}) can be excluded because of the genericity condition as in the proof of Lemma \ref{AnalyticityI} and then, since the coefficients of the equations above are analytic  in a neighborhood of \( p_0 \) and belong to \( \ell_2(\mathbb{Z} \setminus \{n_0 \}) \), \( \{ u_n \}, \{ v_n \} \in \ell_2(\mathbb{Z} \setminus \{n_0 \}) \) are analytic in the same neighborhood.
		\newline
		Moreover \( q_{n_0} \) satisfies the equation
		\bdm
			\bigg\{ 4 \pi \alpha_0 + c_{n_0}(p) + 4 \pi \underset{k \neq n_0}{\sum_{k \in \mathbb{Z}}} \alpha_{k-n_0} v_k \bigg\} q_{n_0} = - 4 \pi \underset{k \neq n_0}{\sum_{k \in \mathbb{Z}}} \alpha_{k-n_0} u_k + f^{(2)}_{n_0}(p)
		\edm
		It is then sufficient to show that 
		\bdm
			\underset{k \neq n_0}{\sum_{k \in \mathbb{Z}}} \alpha_{k-n_0} v_k(p_0) \neq 0
		\edm 
		Let us suppose that the contrary is true: calling \( V_n \equiv v_n(p_0) \), multiplying equation (\ref{Eqvn}) at \( p = p_0 \) by \( V_n^* \) and summing over \( n \in \mathbb{Z} \), \( n \neq n_0 \), one has
		\bdm
			\fl
			\underset{n \neq n_0}{\sum_{n \in \mathbb{Z}}} \left\{ \sqrt{\omega n - ip_0} + \frac{e^{-2r\sqrt{\omega n - ip_0}}}{(2 \pi)^3 r^2(1 - \sqrt{\omega n - ip_0})} \right\} | V_{n} |^2 = - 4 \pi \underset{n,k \neq n_0}{\sum_{n,k \in \mathbb{Z}}} V_n^* \alpha_{k-n} V_k
		\edm
		Using condition (\ref{Conditions2}) and the genericity condition (\ref{Genericity}), as in the proof of Lemma \ref{AnalyticityI}, one obtain \( V_n = 0 \), \( \forall n \in \mathbb{Z} \setminus \{ n_0 \} \), but this is impossible since \( V_n \) satisfies equation (\ref{Eqvn}).
		\newline
		This concludes the proof of analyticity of \( q_2(p) \) in a neighborhood of \( p = p_0 \). In the same way it is possible to conclude that \( q_2(p) \) is also analytic at \( p = i \).
		\newline
		It remains to study the behavior of \( q_1(p) \) and in particular to analyze \( q_0^{(1)}(p) \) in a neighborhood of \( p = i \), where it may have a pole (see equation (\ref{EqL1})): from (\ref{Eqr2}) one has
		\bdm
			\frac{e^{-2r\sqrt{\omega n + 1}}}{(2 \pi)^3 r^2} \: q^{(2)}_n(i) = - \frac{2i \sqrt{2 \pi}}{r} \left[\frac{(2 \pi)^{\frac{3}{2}} - 1}{(2 \pi)^{\frac{3}{2}}} \: e^{-r\sqrt{\omega n + 1}} - e^{-r} \right]
		\edm
		and then \( q_0^{(1)}(i) = i \sqrt{2 \pi} \). \opensquare
	
\textbf{Remark:} Proposition \ref{Poles} holds even if \( \alpha_0 < 0 \). The proof can be given in the same way but it is slightly more complicated, because \( 4 \pi \alpha_0 + c_n(p) \) in Lemma 4.1 could vanish in two points instead of one. Nevertheless the argument contained in Proposition 4.1 can be applied once more, in order to exclude the presence of the corresponding singularity of the solution.

\section{Behavior at \( p = 0 \)}

We shall now study the behavior of the solution of (\ref{EqL1}) and (\ref{EqL2}) in a neighborhood the origin. With the choice (\ref{Branch}) for the branch cut of the square root, it is clear that we must expect branch points of \( \tilde{q}^{(j)}(p) \), solutions of (\ref{Laplace1}) and (\ref{Laplace2}), at \( p = i\omega n \), \( n \in \mathbb{Z} \), which should imply a branch point at \( p = 0 \) for each \( q_n^{(j)} \).
\newline
We are going to show that the solutions of (\ref{EqL1}) and (\ref{EqL2}) have a branch point singularity at the origin.

	\begin{pro}
	\label{BranchPoints}
		\mbox{}	\\
		If \( \{ \alpha_n \} \) satisfies (\ref{Conditions2}) and (\ref{Genericity}) (genericity condition), the solution of the system (\ref{Laplace1}), (\ref{Laplace2}) has the form \( \tilde{q}^{(j)}(p) = c_j(p) + d_j(p) \sqrt{p} \), \( j = 1,2 \), in an imaginary neighborhood of \( p = 0 \), where the functions \( c_j(p) \) and \( d_j(p) \) are analytic at \( p = 0 \).
	\end{pro}

	\emph{Proof:}
		The resonant case, namely if, for some \( N \in \mathbb{N} \), \( \omega = 1/N \), and the non-resonant one will be treated separately.	
		\newline
		\emph{1) Non-resonant case}

		Setting \( q_n = u_n + v_n q_0 \), \( n \neq 0 \) in (\ref{Eqr2}), one obtains the following equations for \( \{ u_n \} \), \( \{ v_n \} \in \ell_2(\mathbb{Z} \setminus \{0\}) \),
		\beq
		\label{Equn0}
			c_n(p) \: u_n = - 4 \pi \underset{k \neq 0}{\sum_{k \in \mathbb{Z}}} \alpha_{k-n} u_k + g_n^{(2)}(p)
		\eeq
		\beq
		\label{Eqvn0}
			c_n(p) \: v_{n} =  - 4 \pi \underset{k \neq 0}{\sum_{k \in \mathbb{Z}}} \alpha_{k-n} v_k - 4 \pi \alpha_{-n}
		\eeq
		If, for every \( n \in \mathbb{Z} \), \( c_n(0) \neq - 4 \pi \alpha_0 \), using the genericity condition, it is easy to prove that \( \{ u_n \} \), \( \{ v_n \} \in \ell_2(\mathbb{Z} \setminus \{0\}) \) are unique and analytic at \( p = 0 \). On the other hand if the condition above is not satisfied and there exists \( N_1 \in \mathbb{Z} \) such that
		\bdm
			4 \pi \alpha_0 + \sqrt{\omega N_1} + \frac{e^{-2r\sqrt{\omega N_1}}}{(2 \pi)^3 r^2(1 - \sqrt{\omega N_1})} = 0
		\edm
		one can repeat the trick, setting for example \( v_n = u_n^{\prime} + v_n^{\prime} v_{N_1} \) for \( n \neq N_1 \), and prove that in fact \( \{ u_n \} \) and \( \{ v_n \} \) are still analytic in a neighborhood of \( p = 0 \).
		\newline
		Thus it is sufficient to prove that \( q_0 \), which is solution of 
		\bdm
			\bigg\{ 4 \pi \alpha_0 + c_0(p) + 4 \pi \: \underset{k \neq 0}{\sum_{k \in \mathbb{Z}}} \alpha_k v_k \bigg\} \: q_0(p) =  - 4 \pi \underset{k \neq 0}{\sum_{k \in \mathbb{Z}}} \alpha_k u_{k} + f_0^{(2)}(p)
		\edm
		has the required behavior near \( p = 0 \). First, setting \( v_n^0 = v_n(p = 0) \), we have to prove that
		\bdm
			\underset{k \neq 0}{\sum_{k \in \mathbb{Z}}} \alpha_k v^0_k \neq - \alpha_0 - \frac{1}{4 \pi (2 \pi)^3 r^2}
		\edm
		but, assuming that the contrary is true and  multiplying both sides of equation (\ref{Eqvn0}), with \( n_0 = 0 \),  by \( {v_n^0}^* \) and summing over \( n \in \mathbb{Z} \), \( n \neq 0 \), one has
		\bdm
			 \underset{n \neq 0}{\sum_{n \in \mathbb{Z}}} \sqrt{\omega n} \: |v^0_n|^2 = - 4 \pi \underset{n,k \neq 0}{\sum_{n,k \in \mathbb{Z}}} {v_n^0}^* \alpha_{k-n} v^0_{k} + 4 \pi \alpha_0 + \frac{1}{(2 \pi)^3 r^2}
		\edm
		The right hand side is still real so that, assuming that the genericity condition is satisfied by \( \{ \alpha_n \} \) and applying the argument contained in the proof of Proposition \ref{Poles}, we immediately obtain \( \{ v^0_n \} = 0 \), which is a contradiction, since \( \{ v_n^0 \} \) solves (\ref{Eqvn0}).
		\newline
		The result for \( \tilde{q}^{(2)} \) follows then directly from the equation for \( q_0 \), since \( e^{-2r\sqrt{-ip}} \) has a branch cut along the negative real line. The extension to \( q^{(1)} \) is thus trivial.	
		\newline
		\emph{2) Resonant case}
		
		As before let us look for a solution of (\ref{Eqr2}) of the form \( q_n = u_n + v_n q_0 \) , \( n \neq 0 \), so that \( \{ u_n \} \), \( \{ v_n \} \in \ell_2(\mathbb{Z} \setminus \{0\}) \) solve (\ref{Equn0}) and (\ref{Eqvn0}) with \( \omega = 1/N \). Multiplying both sides of (\ref{Equn0}) and (\ref{Eqvn0}) for \( n = N \) by \( 1 - n/N - ip \), one sees that \( u_N \) and \( v_N \) have no pole singularity at \( p = 0 \). On the other hand, if there exists \( N_1 \in \mathbb{Z} \) such that
		\bdm
			4 \pi \alpha_0 + \sqrt{\frac{N_1}{N}} + \frac{e^{-2r\sqrt{N_1/N}}}{(2 \pi)^3 r^2(1 - \sqrt{N_1/N})} = 0
		\edm
		the solutions could have a pole at \( p = 0 \), for \( n = N_1 \) (the expression above guarantees that \( N_1 \neq N \)). Nevertheless, repeating the above procedure for \( n = N_1 \), it is easily seen that in fact \( \{ u_n \} \), \( \{ v_n \} \in \ell_2(\mathbb{Z} \setminus \{0\}) \) are both analytic in a neighborhood of \( p = 0 \).
		\newline
		The behavior of \( q^{(2)} \) near \( p = 0 \) is then proved like in the non-resonant case, but we have now to take care about \( q^{(1)} \), since the coefficient in \( \mathcal{M}_1 \) for \( n = N \) (see the definition (\ref{M1Operator})) has a pole at \( p = 0 \). But from (\ref{Eqr2}) one has
		\bdm
			\frac{e^{-2r\sqrt{n/N - ip}}}{(2 \pi)^3 r^2} \:\: q^{(2)}_n(0) = - \frac{2i \sqrt{2 \pi}}{r (1+\sqrt{n/N})} \left[ \frac{(2 \pi)^{\frac{3}{2}} - 1}{(2 \pi)^{\frac{3}{2}}}  \: e^{-r\sqrt{n/N}} - e^{-r} \right]
		\edm
		so that \( q_N^{(1)}(0) = i \sqrt{2\pi} \). \opensquare
 
\section{Complete ionization in the generic case}

Summing up the results about the behavior of the Laplace transforms \( \tilde{q}^{(j)}(p) \), \( j = 1,2 \), we can state the following
	
	\begin{teo}
	\label{Decay}
		If \( \{ \alpha_n \} \) satisfies (\ref{Conditions2}) and the genericity condition (\ref{Genericity}) with respect to \( \mathcal{T} \), as \( t \rightarrow \infty \),
		\beq
			|q^{(j)}(t)| \leq A_j \:  t^{-\frac{3}{2}} + R_j(t)
		\eeq
		where \( A_j > 0 \) and \( R_j(t) \) has an exponential decay, \( R_j(t) \sim C_j e^{-B_jt} \) for some \( B_j > 0 \).
		\newline
		Moreover the system shows asymptotic complete ionization and, as \( t \rightarrow \infty \), 
		\bdm
			| \theta(t) | = \Big| \Big( \varphi_{\alpha(0)} \: , \Psi_t \Big) \Big| \leq D \: t^{-\frac{3}{2}} + E(t)
		\edm
		where \( D > 0 \) and \( E(t) \) has an exponential decay.
	\end{teo}

	\emph{Proof:}	Propositions \ref{Analyticity}, \ref{Poles} and \ref{BranchPoints} guarantee that \( \tilde{q}(p) \) is analytic on the closed right half plane, except branch point singularities on the imaginary axis at \( p = i \omega n \), \( n \in \mathbb{Z} \). Therefore we can chose a integration path for the inverse of Laplace transform of \( \tilde{q}(q) \) along the imaginary axis like in \cite{Cost1} and the result is a straightforward consequence of the behavior of \( q^{(j)}(p) \) around the branch points given by Proposition \ref{BranchPoints} (see e.g. the proof of Theorem 3.1 in \cite{Corr2}).
		\newline
		The Laplace transform of \( \theta(t) \) can be expressed in the following way (see e.g. Proposition 2.1 in \cite{Corr2})
		\bdm
			\tilde{\theta}(p) = \tilde{Z}(p) + \tilde{Z}_1(p) \: \tilde{q}^{(1)}(p) + \tilde{Z}_2(p) \: \tilde{q}^{(2)}(p)
		\edm 
		where \( \tilde{Z}(p) \) is analytic on the closed right half plane and \( \tilde{Z}_j(p) \) has only a branch point at the origin of the form \( a_j + b_j \sqrt{p} \). Hence \( \tilde{\theta}(p) \) has the same singularities as \( \tilde{q}(p) \) and then its asymptotic behavior coincides with that of \( q(t) \). \opensquare

In the following we shall prove a stronger result about complete ionization of the system, namely that every state \( \Psi \in L^2(\mathbb{R}^3) \) is a scattering state for the operator \( H(t) \), i.e. 
	\beq
	\label{Ioni}
		\lim_{t \rightarrow \infty} \frac{1}{t} \int_0^t d\tau \: \big\| F(|\bi{x}| \leq R) U(\tau,0) \Psi \big\|^2 = 0
	\eeq
where \( F(S) \) is the multiplication operator by the characteristic function of the set \( S \subset \mathbb{R}^3 \) and \( U(t,s) \) the unitary two-parameters family associated to \( H(t) \) (see (\ref{Schro})).
\newline
In order to prove (\ref{Ioni}), we first need to study the evolution of a generic initial datum in a suitable dense subset of \( L^2(\mathbb{R}^3) \) and then we shall extend the result to every state using the unitarity of the evolution defined by (\ref{Schro}) (see e.g. \cite{Dell1}).
	
	\begin{pro}
	\label{DecayG}
		Let \( \Psi \in C^{\infty}_0(\mathbb{R}^3 \setminus \{ 0, \bi{r} \}) \) a smooth function with compact support away from \( 0, \bi{r} \) and \( q^{(j)}(t) \) be the solutions of equations (\ref{Equation1}) and (\ref{Equation2}) with initial condition \( \Psi_0 = \Psi \). If \( \{ \alpha_n \} \) satisfies (\ref{Conditions2}) and the genericity condition (\ref{Genericity}) with respect to \( \mathcal{T} \), as \( t \rightarrow \infty \),
		\beq
			| q^{(j)}(t) | \leq A_j \:  t^{-\frac{3}{2}} + R_j(t)
		\eeq
		where \( A_j > 0 \) and \( R_j(t) \) has an exponential decay, \( R_j(t) \sim C_j e^{-B_j t} \) for some \( B_j > 0 \).
	\end{pro}

	\emph{Proof:}	
		The estimate on the behavior for large time contained in Section 2 still applies, so that \( \tilde{q}^{(j)}(p) \) is analytic \( \forall p \) with \( \Re(p) > b_0 \).
		\newline
		Hence we can consider the Laplace transforms of equations (\ref{Equation1}) and (\ref{Equation2}), which have the form (\ref{EqL1}) and (\ref{EqL2}) with
		\bdm
			G_1(p) = \sqrt{\frac{2}{\pi}} \: \int_0^{\infty} dt \: e^{-pt} \int_{\mathbb{R}^3} d^3 \bi{k} \:\: \hat{\Psi}(\bi{k}) \: e^{-ik^2t}
		\edm
		\bdm
			G_2(p) = \sqrt{\frac{2}{\pi}} \: \int_0^{\infty} dt \: e^{-pt} \int_{\mathbb{R}^3} d^3 \bi{k} \:\: \hat{\Psi}(\bi{k}) \: e^{-i (k^2-\bi{k} \cdot \bi{r} ) \: t}
		\edm
		where \( \hat{\Psi}(\bi{k}) \) is the Fourier transform of \( \Psi \).
		\newline
		Since for every smooth function \( \Psi \) with compact support, \( \hat{\Psi}(\bi{k}) \) is a smooth function with an exponential decay as \( k \rightarrow \infty \), so that \( G_j(p) \) has the same singularities as in the case already studied, i.e. a branch point at the origin of the form \( a(p) + b(p) \sqrt{p} \). \opensquare

	\begin{teo}
	\label{Scattering}
		If \( \{ \alpha_n \} \) satisfies (\ref{Conditions2}) and the genericity condition (\ref{Genericity}) with respect to \( \mathcal{T} \), every \( \Psi \in L^2(\mathbb{R}^3) \) is a scattering state of \( H(t) \), i.e.
		\bdm
			\lim_{t \rightarrow \infty} \frac{1}{t} \int_0^t d\tau \: \big\| F(|\bi{x}| \leq R) U(\tau,0) \Psi \big\|^2 = 0
		\edm
		Moreover the discrete spectrum of the Floquet operator associated to \( H(t) \),
		\bdm
			K \equiv -i \frac{\partial}{\partial t} + H(t)
		\edm
		is empty.
	\end{teo}
		
	\emph{Proof:}
		The proof follows from unitarity of the evolution and the explicit expression (\ref{State}), together with Proposition \ref{DecayG} (see the proof of Theorem 3.2 in \cite{Corr2}). The absence of eigenvalues of the Floquet operator is a straightforward consequence: every eigenvector of \( K \) is of the form \( e^{i \beta t} \chi(\bi{x},t) \), where \( \beta \in \mathbb{R} \) and \( \chi \) is periodic in time, hence it can not satisfy (\ref{Ioni}). \opensquare

\end{document}